\documentclass{article}
\usepackage[title]{appendix}
\usepackage{times}

\usepackage[dvips]{graphicx}
\usepackage{epsfig}
\usepackage{hyperref}
\usepackage{latexsym}
\usepackage{xspace}
\usepackage{subcaption}

\usepackage{amsmath}

\usepackage{amsthm}
\usepackage{amssymb}
\usepackage{mathtools}
\usepackage{color}
\usepackage{algorithm}
\usepackage{algpseudocode}

\algnewcommand{\Inputs}[1]{%
  \State \textbf{Inputs:}
  \Statex \hspace*{\algorithmicindent}\parbox[t]{.8\linewidth}{\raggedright #1}
}
\algnewcommand{\Outputs}[1]{%
  \State \textbf{Outputs:}
  \Statex \hspace*{\algorithmicindent}\parbox[t]{.8\linewidth}{\raggedright #1}
}
\algnewcommand{\Initialize}[1]{%
  \State \textbf{Initialize:}
  \Statex \hspace*{\algorithmicindent}\parbox[t]{.8\linewidth}{\raggedright #1}
}

\DeclareMathOperator{\Beta}{Beta}

\theoremstyle{definition}

\theoremstyle{plain}

\begin{document}

\title{Bayesian Beta-Binomial Prevalence Estimation Using an Imperfect
  Test} \author{Jonathan Baxter\\\tt{covid@baxters.biz}} \date{April 23, 2020}\maketitle
\abstract{Following \cite{diggle_2011,greenland_1996}, we give a
  simple formula for the Bayesian posterior density of a prevalence
  parameter based on unreliable testing of a population. This problem
  is of particular importance when the false positive test rate is
  close to the prevalence in the population being tested. An efficient
  Monte Carlo algorithm for approximating the posterior density is
  presented, and applied to estimating the Covid-19 infection rate in
  Santa Clara county, CA using the data reported in
  \cite{Bendavid2020.04.14.20062463}. We show that the true Bayesian
  posterior places considerably more mass near zero, resulting in a
  prevalence estimate of 5,000--70,000 infections (median: 42,000)
  (2.17\% (95CI 0.27\%--3.63\%)), compared to the estimate of
  48,000--81,000 infections derived in
  \cite{Bendavid2020.04.14.20062463} using the delta method.

  A demonstration, with code and additional examples, is available at
  \url{https://testprev.com}.  }

\section{Introduction}
We consider the problem of estimating disease prevalence in a
population of interest using an unreliable test. Following
\cite{diggle_2011,greenland_1996}, we take a Bayesian approach and
model uncertainty in the test characteristics (sensitivity and
specificity) as well as the uncertainty due to a finite testing
population. We extend \cite{diggle_2011} by deriving a simple
expression for the posterior prevalence probabilty density in the
common case of a test that has been validated against a number of
known positive and negative subjects. A Monte Carlo algorithm for
computing the prevalence posterior is presented, and applied to
Covid-19 infection data from Santa Clara county
CA~\cite{Bendavid2020.04.14.20062463}, where the false positive test
calibration rate (0.5\%) was close to the measured prevalence
(1.5\%). The posterior distribution on prevalence in this case
acquires a second mode at zero, which results in substantial
broadening of the credible interval on prevalence. The appearance of a
second mode also explains why local approximation methods such as the
delta-method can fail to capture all the posterior variance.

\section{Known test performance}
Suppose we know the test false-positive rate (1 - specificity) $u$ and
sensitivity $v$. Denote the (unknown) population prevalence by
$\theta$. The probability $p$ of a positive test is the probability of
a positive test given the subject has the disease, plus the
probability of a positive test given the subject is disease-free:
\begin{equation}
  \label{eq:p}
  p = v\theta + u(1-\theta) = u + \theta(v-u)
\end{equation}
The probability of $k$ positive tests out of $n$ subjects tested
follows a binomial distribution with parameter $p$:
\begin{equation}
  \label{eq:bin_p}
  \Pr\left(k|n,\theta,u,v\right) = {n \choose k} p^k(1-p)^{n-k}
\end{equation}
By application of Bayes' rule, the distribution of $\theta$ is given by:
\begin{equation}
  \label{eq:p_theta}
  \Pr\left(\theta|k,n,u,v\right) = \frac{\Pr\left(k|n,\theta,u,v\right)\Pr(\theta)}{\Pr\left(k|n,u,v\right)}
\end{equation}
where $\Pr(\theta)$ is our prior probability on $\theta$ and
\begin{equation}
  \label{eq:p_k}
  \Pr\left(k|n,u,v\right) = \int_0^1\Pr\left(k|n,\theta,u,v\right)\Pr(\theta)\,d\theta
\end{equation}
Choosing a uniform prior on $\theta$, and applying $d\theta = \frac{dp}{v-u}$:
\begin{align}
\Pr\left(k|n,u,v\right) \nonumber
&= \int_0^1 {n \choose k} p^k(1-p)^{n-k}\,d\theta \\\nonumber
   &= \frac{1}{v-u}{n \choose k}\int_u^v p^k(1-p)^{n-k}\,dp \\\nonumber
   &= \frac{1}{v-u}{n \choose k}\left[\int_0^v p^k(1-p)^{n-k}\,dp - \int_0^u p^k(1-p)^{n-k}\,dp\right] \\\nonumber
   &= \frac{1}{v-u}{n \choose k}\left[B(v;k+1,n-k+1) - B(u;k+1,n-k+1)\right] \\
&=:\frac{B(v) - B(u)}{v-u}{n \choose k}\label{eq:k_prior}
\end{align}
where $B(x;\alpha,\beta) := \int_0^x t^{\alpha-1}(1-t)^{\beta-1}\, dt$
is the incomplete Beta function, and for notational brevity we drop
the dependence on $k$ and $n$ from $B(v;k+1,n-k+1)$ and just write
$B(v)$.

Substituting~\eqref{eq:p}, ~\eqref{eq:bin_p} and \eqref{eq:k_prior} into \eqref{eq:p_theta} yields:
\begin{equation}
  \label{eq:theta_known}
  \Pr\left(\theta|k,n,u,v\right) = \frac{v-u}{B(v)-B(u)} [u + \theta(v-u)]^k [1- u - \theta(v-u)]^{n-k}
\end{equation}

\section{Estimated test performance}
Equation ~\eqref{eq:theta_known} expresses the distribution over
population prevalence $\theta$ given known test characteristics $u$
and $v$. However, the false-positive rate $u$ and sensitivity $v$ are
usually themselves estimates based on validation against known
positive and negative subjects. Specifically, suppose the test has
been validated with $k_u$ false positives out of $n_u$ known negative
samples, and $k_v$ true positives out of $n_v$ known positive samples.

Assuming a beta prior on $u$ with parameters $\alpha_u, \beta_u$, the
posterior density on $u$ given the validation data is proportional to
a beta density with parameters $k_u+\alpha_u,n_u-k_u+\beta_u$. Abusing
notation for clarity, write $\Beta_u(u)$ for this density and similarly
$\Beta_v(v)$ for the corresponding density on $v$. Let $\Beta_p(u
+\theta(v-u))$ denote the density at $u +\theta(v-u)$ of the beta
distribution with parameters $k+1,n-k+1$.

With this notation, integrating out $u,v$ from \eqref{eq:theta_known},
we obtain the following expression for the posterior distribution on
prevalence $\theta$ that accounts for uncertainty in the test
characteristics:
\begin{multline}
  \label{eq:theta_unknown}
  \Pr\left(\theta|k,n,k_u,n_u,k_v,n_v\right) \propto \\
  \int_0^1\int_0^v\frac{v-u}{B(v)-B(u)} \Beta_p(u + \theta(v-u))\Beta_u(u)\Beta_v(v)\,du\,dv
\end{multline}
The domain of integration has been restricted to the region 
$v-u>0$, reflecting the fact that a test with false-positive rate $u$
in excess of sensitivity $v$ is not a usable test (this can also be
thought of as an adjustment of the joint posterior on $u$ and $v$ to
capture a dependence between $u$ and $v$).

To the author's knowledge there is no closed-form expression for the
right-hand-side of \eqref{eq:theta_unknown} in terms of hypergeometric
or related functions. In the next section we will describe an
algorithm for evaluating the integral using Monte Carlo integration.

\section{Computing the Posterior Distribution}
Let $I(\theta)$ denote the integral on the right-hand-side of
\eqref{eq:theta_unknown}.  Draw $N$ samples $u_i,v_i$ from
$\Beta_u(u)$ and $\Beta_v(v)$ (any pairs such that $u_i>v_i$ are
rejected and resampled). Then with error $\sim \frac{1}{\sqrt{N}}$,
\begin{equation}
  \label{eq:approx}
  I(\theta) \approx \frac1N\sum_{i=1}^N \frac{v_i-u_i}{B(v_i)-B(u_i)}
  \Beta_p(u_i + \theta(v_i-u_i))
\end{equation}
This expression is calculated for a discrete grid of values $\theta_j$
and then normalized to generate the posterior density $\Pr(\theta)$.
The samples $u_i,v_i$ can be reused for each estimate $I(\theta_j)$,
which allows computation of $\frac{v_i-u_i}{B(v_i)-B(u_i)}$ to be
performed once and reused.

To avoid numerical problems in the case $B(u) \approx B(v) \approx 1$,
observe that $B(u;k+1,n-k+1) = B(k+1,n-k+1)-B(1-u;n-k+1,k+1)$ where
$B(k+1,n-k+1)$ is the complete beta function with parameters $k+1,
n-k+1$. Thus
\begin{equation}
  \label{eq:bsub}
  B(v) - B(u) = B(1-u; n-k+1,k+1) - B(1-v; n-k+1,k+1).
\end{equation}
The right-hand-side of \eqref{eq:bsub} is the difference of two values
close to zero when the left-hand-side is the difference of two values
close to 1. Differencing two small numbers has better numerical
stability than differencing two numbers that may be indistinguishable
from 1 within machine precision.

With this substitution, Algorithm \ref{alg:PPP} gives pseudocode for
computing the full prevalance posterior given the results of an
imperfect test.

\begin{algorithm}[H]
  \caption{Posterior prevalence probability (PPP) estimation from an imperfect test}
  \label{alg:PPP}
    \begin{algorithmic}[1]
      \Inputs{$k,n,k_u,n_u,k_v,n_v,\alpha_u,\beta_u,\alpha_v,\beta_v,N,M$\\
        $B(\cdot)$: incomplete beta function with parameters $n-k+1, k+1$}
      \Outputs{Posterior prevalence probability density $p_j$ at $\frac{j}{M}$, $j=0\dots M$} \\ \\
      {\bf Initialization:}
      \For{i=1 to N}
      \State $u_i\gets 0, v_i\gets 0$
      \While{$u_i \geq v_i$}
      \State $u_i \gets u \sim \Beta(k_u + \alpha_u, n_u-k_u+\beta_u)$
      \State $v_i \gets v \sim \Beta(k_v + \alpha_v, n_v-k_v+\beta_v)$
      \EndWhile
      \State $d_i \gets \frac{v_i-u_i}{B(1-u_i) - B(1-v_i)}$
      \EndFor \\\\
      {\bf Posterior Density Estimation:}
      \For{j=0 to M}
      \State $\theta_j\gets \frac{j}{M}, p_j\gets 0$
      \For{i=1 to N}
      \State sample $f \sim \Beta(u_i + \theta_j(v_i-u_i); k+1, n-k+1)$
      \State $p_j \gets p_j + d_i*f$
      \EndFor
      \State $p_j \gets \frac{p_j}{N}$
      \EndFor \\\\
      {\bf Normalization:}
      \State $T \gets \frac1{M+1}\sum_{j=0}^M p_j$
      \State $p_j\gets \frac{p_j}{T}, j=0,\dots,M$
    \end{algorithmic}
\end{algorithm}

\section{Example}
The prevalence of SARS-CoV-2 antibodies in Santa Clara county, CA, was
recently measured using an imperfect serological test
\cite{Bendavid2020.04.14.20062463}. Three different calculations were
performed based on different estimates of the test's sensitivity and
specificity. For brevity, we will focus on their scenario 3 (similar
conclusions apply to the other two scenarios).

The relevant parameters for the PPP estimation algorithm are as
follows:

\begin{itemize}
\item  $k=50$ positive tests out of $n=3330$ subjects tested.
\item $k_u=2$ false positives out of $n_u=401$ known negative samples.
\item $k_v=103$ correct positives out of $n_v=122$ known positive samples.
\end{itemize}

The authors used the delta method~\cite{delta_1992} to estimate
standard errors for the population prevalence, which accounts for
sampling error and propagates the uncertainty in the test sensitivity
and specificity. However, the delta method provides only a local
approximation to the posterior density, and with small counts this can
result in underestimated variance.

The raw positive test count $k=50$ was also reweighted to account for
demographic differences between the test sample and the overall Santa
Clara population, yielding a considerably larger population-adjusted
count of $k=94$.  The reweighting was applied before the
sensitivity/specificity adjustments, which also has potential to
underestimate variance in the final result.

After all adjustments, the authors reported a prevalence estimate of
2.75\% (95CI 2.01\%--3.49\%).

In order to avoid potentially biasing our results, we applied the PPP
algorithm to the raw counts, and then the population reweighting was
applied to the estimated posterior prevalence distribution. To compare
with the delta method used in~\cite{Bendavid2020.04.14.20062463}, we
reran their methodology adjusting first for uncertainty in the test
characteristics, and then for population. Omitting the details, we
arrive at a prevalence estimate of 2.81\% (95CI 1.74\% --
3.88\%). Observe that while the lower bound of the CI has dropped from
2.01\% to 1.74\%, it is still well above zero.

\begin{figure}
  \begin{subfigure}{0.5\textwidth}
    \centering
    \includegraphics[width=0.97\linewidth]{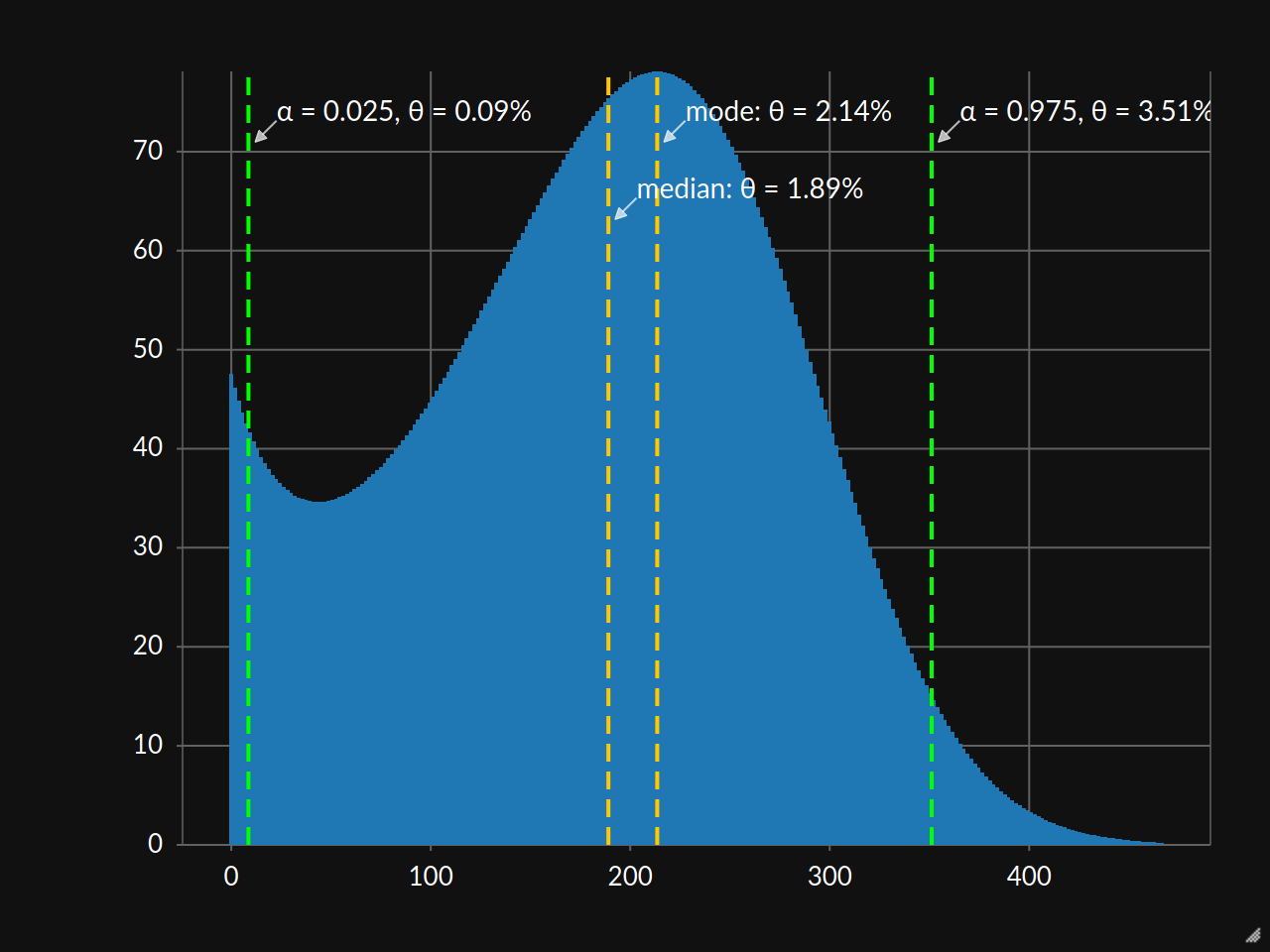}
    \caption{}
    \label{fig:sc_ppp}
  \end{subfigure}%
  \begin{subfigure}{0.5\textwidth}
    \centering
    \includegraphics[width=0.97\linewidth]{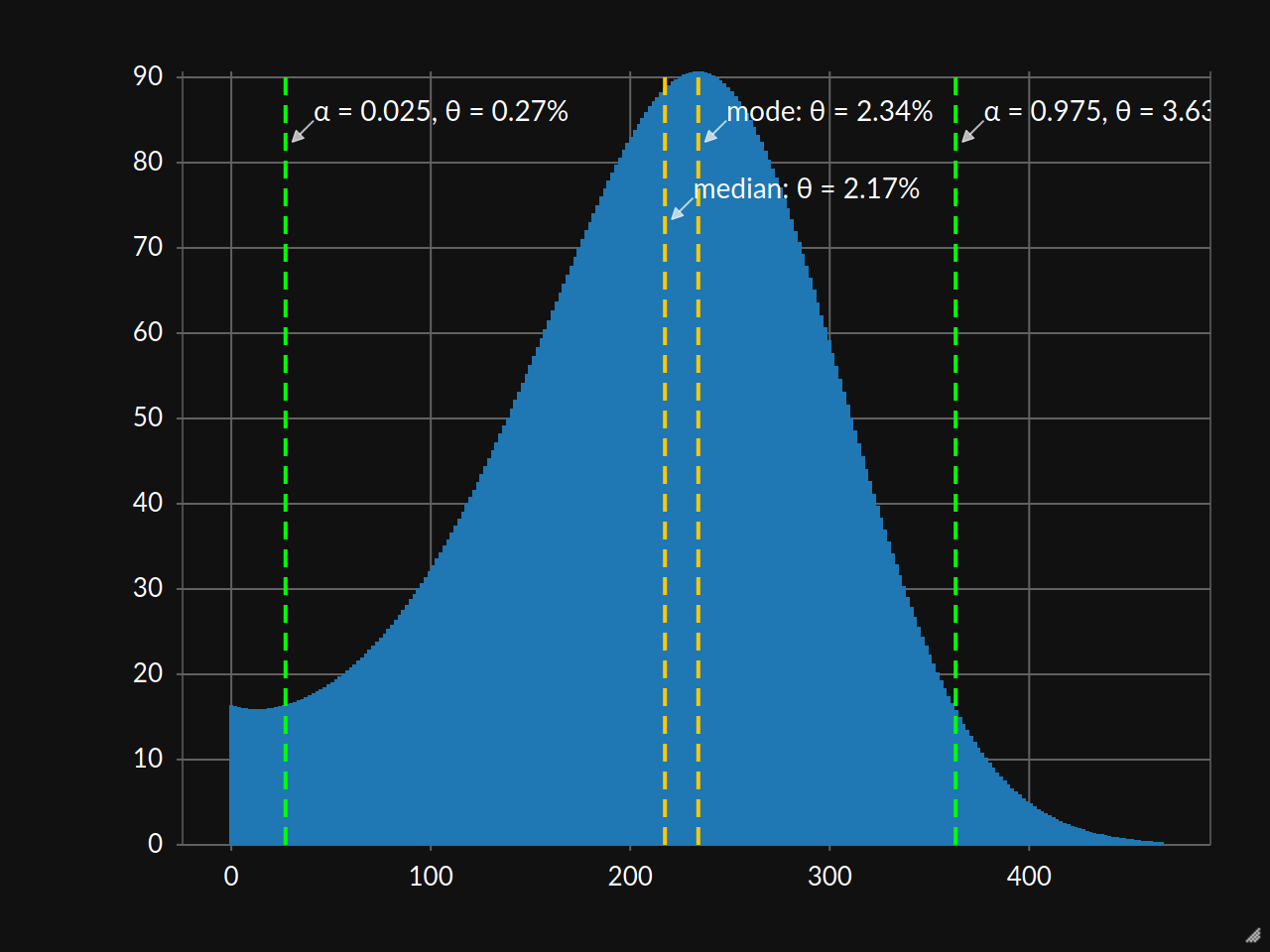}
    \caption{}
    \label{fig:sc_inform}
  \end{subfigure}
  \caption{Prevalence posterior for Santa Clara county testing data
    described in ~\cite{Bendavid2020.04.14.20062463}. The $x$-axis is
    measured in bps, or units of 0.01\%. Figure~\ref{fig:sc_ppp}:
    uniform priors on test false positive rate and sensitivity.
    Figure~\ref{fig:sc_inform}: $\Beta(1,99)$ prior on test false
    positive.}
  \label{fig:posterior}
\end{figure}

Figure~\ref{fig:sc_ppp} shows the prevalence posterior density
computed using the PPP algorithm with uniform priors
($\alpha_u=\beta_u=\alpha_v=\beta_v=1$), a Monte Carlo sample size $N$
of 10,000, and a grid size $M$ of 10,000. The estimated prevalence of
Covid-19 in Santa Clara county is 1.89\% (95CI 0.09\% -- 3.51\%), with
a notably reduced lower bound on the credible interval of 0.09\%. This
translates to an infected population range of 1,800--68,000,
considerably wider than the 38,000--76,000 range derived using the
delta method in ~\cite{Bendavid2020.04.14.20062463}.

The shape of the prevalence posterior near zero helps explain the
difference between the two results. The full Bayes approach assigns
considerably more mass towards zero, such that the posterior
distribution becomes bimodal. This is driven by two factors:
\begin{itemize}
  \item The uncertainty in the false positive rate $u$, which is
    derived from only $k_u=2$ false positves out of $n_u=401$ known
    negative samples (0.5\%).
  \item The underlying infection prevalence rate (estimated at 1.5\%)
    is close to the test false positive rate (0.5\%).
\end{itemize}

The uncertainty in false positive rate is exacerbated by using a
uniform prior on $u$, which is arguably too conservative in this
case. Figure~\ref{fig:sc_inform} shows the posterior generated with
$\alpha_u=1, \beta_u=99$, which corresponds to a beta prior with mean
and standard deviation of 1\%. Note that the bimodality is almost
eliminated, but the credible interval is still considerably wider than
that derived via the delta method: 2.17\% (95CI 0.27\%--3.63\%). This
corresponds to an infected range of 5,000--70,000 with median 42,000.

\section{Discussion}
Reliably estimating infection prevalence with an unreliable diagnostic
test is of particular importance during the Covid-19 pandemic,
expecially when the infection prevalence is not much greater than the
test's false positive rate.  Following
\cite{diggle_2011,greenland_1996}, we derived a simple expression
\eqref{eq:theta_unknown} for the posterior prevalence distribution
given the results of an unreliable diagnostic test. A Monte Carlo
algorithm (Posterior Prevalence Probability or PPP) for efficiently
computing the posterior was given. Application of the algorithm to the
Santa Clara county, CA Covid-19 test data
in~\cite{Bendavid2020.04.14.20062463} generates credible intervals
with considerably more mass at zero than the delta method used in the
same paper. This is primarily due to the appearance of a second mode
in the posterior density at zero, which is not captured by local
methods such as the delta method.

A demonstration (with code and additional examples) is available at
\url{https://testprev.com}.

\subsection*{Acknowledgements}
Thanks to David Joerg and Charlie Graham for helpful comments on an
earlier draft of this paper.

\bibliographystyle{abbrv}
\bibliography{bib}
\end{document}